# Enhancement of anomalous Nernst effect in Ni/Pt superlattice

T. Seki,[1,2,3,*] Y. Sakuraba,[3,4] K. Masuda,[3] A. Miura,[3] M. Tsujikawa,[2,5] K. Uchida,[1,2,3]

T. Kubota,[1,2] Y. Miura,[3] M. Shirai,[2,5,6] and K. Takanashi[1,2,6]

[1] Institute for Materials Research, Tohoku University, Sendai 980-8577, Japan

[2] Center for Spintronics Research Network, Tohoku University, Sendai 980-8577, Japan

[3] National Institute for Materials Science, Tsukuba 305-0047, Japan

[4] PRESTO, Japan Science and Technology Agency, Saitama 322-0012, Japan

[5] Research Institute of Electrical Communication, Tohoku University, Sendai 980-8577, Japan

[6] Center for Science and Innovation in Spintronics, Core Research Cluster, Tohoku University, Sendai

980-8577, Japan

* e-mail: go-sai@imr.tohoku.ac.jp

**Abstract:**

We report an enhancement of the anomalous Nernst effect (ANE) in Ni/Pt (001) epitaxial superlattices. The transport and magneto-thermoelectric properties were investigated for the Ni/Pt superlattices with various Ni layer thicknesses ($t$). The anomalous Nernst coefficient was increased up to more than 1 μV K$^{-1}$ for 2.0 nm $\leq t \leq$ 4.0 nm, which was the remarkable enhancement compared to the bulk Ni. It has been found that the large transverse Peltier coefficient ($\alpha_{xy}$), reaching $\alpha_{xy}$ = 4.8 A K$^{-1}$ m$^{-1}$ for $t$ = 4.0 nm, plays a prime role for the enhanced ANE of the Ni/Pt (001) superlattices.

Spin caloritronics [Ref.1], the field studying the interconversion between charge current ($\mathbf{J}_c$) and heat current ($\mathbf{J}_q$) mediated by spin current ($\mathbf{J}_s$) and/or magnetization ($\mathbf{M}$), has attracted much attention not only for academic interests but also for practical applications. The newly discovered spin caloritronic phenomena such as spin Seebeck effect [Refs.2-4] have stimulated the renewed interest in the well-known thermoelectric phenomena in ferromagnets. One of the thermoelectric phenomena in ferromagnets is the anomalous Nernst effect (ANE), in which $\mathbf{J}_c$ appears in the cross-product direction of $\mathbf{M}$ and a temperature gradient ($\nabla T$). Although ANE has been known for a long time, the microscopic physical picture for ANE has not fully been understood yet. In addition to the fundamental point of view, this magneto-thermoelectric effect is possibly beneficial for thermoelectric conversion applications [Refs.5,6]. The key for the ANE-based thermoelectric conversion is to find a material with a large anomalous Nernst coefficient ($S^{ANE}$) because the charge current density induced by ANE ($\mathbf{j}_{c,ANE}$) is given by $\mathbf{j}_{c,ANE} = \sigma S^{ANE}\{(\mathbf{M}/|\mathbf{M}|) \times \nabla T\}$ with electrical conductivity ($\sigma$) [Ref.7].

Several ferromagnets show the ANE and the anomalous Ettingshausen effect as the reciprocal phenomenon [Refs.8-16]. We previously reported the enhancement of ANE in the metallic multilayers of Fe/Pt, Fe/Au, and Fe/Cu [Ref.17]. Fang *et al.* [Ref.18] also reported that ANE was increased with the number of interfaces for the Co/Pt superlattices. These reports imply the low dimensionality of layer and/or the existence of interface plays a crucial role for the increase in ANE, and one may be aware that metallic multilayers or superlattices with a number of interfaces are possible candidates for achieving large ANE.

Among several choices for a ferromagnet and a paramagnet composed of the metallic multilayer or superlattice, this study focuses on ferromagnetic Ni and paramagnetic Pt. Ni is a material exhibiting the large anisotropic magneto-Peltier effect thanks to it characteristic electronic structure [Refs.19,20], and is an interesting material from the viewpoint of ANE [Ref.21]. On the other hand, Pt is a representative paramagnet having the large spin-orbit interaction. This large spin-orbit interaction of Pt is probably advantageous for the ANE of Ni through the interface. Recently, we successfully prepared the perpendicularly magnetized Ni/Pt (001) epitaxial superlattices on a non-conductive $SrTiO_3$ substrate without any buffer layer materials [Ref.22]. The Ni/Pt (001) epitaxial superlattices are suitable for studying the effects of layer thickness and interface on the magnitude of ANE, and available to compare the experimental results with theoretical calculation. In this paper, we report the investigation of ANE in the Ni/Pt (001) epitaxial superlattices with various Ni layer thicknesses ($t$). In addition to the evaluation of $S^{ANE}$, the value of $S^{ANE}$ divided by saturation magnetization ($M_s$) is shown, which is an indicator for the ANE-based thermopiles integrated densely [Ref.23]. We found the enhanced ANE of the Ni/Pt (001) superlattices, which is attributable to the large transverse Peltier coefficient ($\alpha_{xy}$).

[Ni ($t$)/Pt (1.0 nm)]$_{\times N}$ supelattices were grown on SrTiO$_3$ (100) single crystal substrates employing a magnetron sputtering system with the base pressure below 2 × 10$^{-7}$ Pa. The deposition temperature was set at 400ºC for the Ni and Pt layers. The Ni layer was first deposited, which was followed by the layers of [Pt/Ni]$_{\times N-1}$/Pt. Finally, a 2 nm-thick Al layer was deposited at room temperature as a capping layer. The substrate temperature of 400ºC was necessary to achieve the (001) epitaxial growth of the Ni/Pt superlattices, and the well-defined layered structures were achieved without remarkable intermixing between the layers as reported in Ref.22. The magnetic properties for the thin films were measured using a vibrating sample magnetometer (VSM) at room temperature. The thin films were patterned into Hall-cross shapes through the use of photolithography and Ar ion milling. This study exploited two different Hall-cross-shaped devices in accordance with the purpose of measurement: one is for the electrical transport measurement, and the other is for the thermoelectric measurement. For evaluating the electrical transport properties, the devices were installed into the physical properties measurement system (PPMS, Quantum Design, Inc.), and the magnetic field dependence of longitudinal ($\rho_{xx}$) and transverse resistivities ($\rho_{xy}$) was measured with varying the measurement temperature ($T$). For the evaluation of the ANE, we gave $\nabla T$ to the in-plane direction and applied external magnetic field to the perpendicular direction to the Hall bar to measure the electric field ($E_{ANE}$) arising from ANE in PPMS. $\nabla T$ in PPMS was carefully estimated using the procedure described in Refs. 13-15 with the infra-red camera. The Seebeck effects as well as $\sigma$ for the blanket films were measured employing the Seebeck coefficient/electric resistance measurement system (ZEM-3, ADVANCE RIKO, Inc.). All the measurements for the thermoelectric properties were carried out at room temperature.

**Figure 1(a)** shows the magnetization curves for the [Ni/Pt]$_{\times N}$ with $t$ = 1.5, 2.0, 3.0 and 4.0 nm, where $N$ was set to be 8, 7, 5, and 4, respectively. Those repetition numbers were adjusted for the total thicknesses of approximately 20 nm. The red curves denote the magnetization curves measured with the magnetic field ($H$) applied in the film plane (IP curve) while the blue curves denote those measured with the out-of-plane magnetic field (OPP curve). In this study, $M$ was defined as the detected magnetic moment per the unit volume of Ni layers. All the films show the perpendicular magnetization. The effective uniaxial magnetic anisotropy constant ($K_{eff}$) corresponds to the area enclosed between the OPP and IP curves. The values of $M_s$ and uniaxial magnetic anisotropy constant ($K_u$) as a function of $t$ are plotted in **Figs. 1(b) and 1(c)**, where $K_u = K_{eff}+2\pi M_s^2$. $M_s$ is decreased as $t$ is reduced, which results from the decrease in Curie temperature at the small $t$ [Ref.22]. $K_u$ also shows the reduction with decreasing $t$. This $t$ dependence of $K_u$ is partially related with that of $M_s$. The other reason is that the adequate thickness region to obtain the large $K_u$ exists for the Ni / Pt (001) superlattice, which is 2.0 nm $\leq t \leq$ 4.0 nm as reported previously [Ref.22].

The electrical transport properties were measured as illustrated in **Fig. 2(a)**, where the width of Hall bar is 10 μm and the edge-to-edge distance between the Hall branches is 50 μm. The longitudinal voltage ($V_{xx}$) and transverse voltage ($V_{xy}$) were measured under the application of dc current ($I_{dc}$) and perpendicular magnetic field ($H_z$). **Figure 2(a)** displays the transverse resistance ($R_{xy}$) for the device with $t$ = 3.0 nm. The square-shaped hysteresis is observed. $\rho_{xy}$ is composed of two terms: ordinary Hall effect and anomalous Hall effect (AHE). For the present perpendicularly magnetized Ni/Pt superlattice, $\rho_{xy}$ at $H_z$ = 0 Oe is the value coming from only the AHE term. **Figures 2(b) and 2(c)** plot the longitudinal conductivity ($\sigma_{xx}$) and the transverse conductivity ($\sigma_{xy}$), respectively, at $H_z$ = 0 Oe as a function of $T$. Regardless of $t$, the metallic behavior is observed in the $T$ dependence of conductivities. **Figure 2(d)** corresponds to the $\sigma_{xx}$ versus $|\sigma_{xy}|$ plot. Onoda *et al.* [Ref.24] mentioned that $\sigma_{xy}$ shows a gradual dependence on $\sigma_{xx}$ and becomes constant of $10^2$ - $10^3$ $\Omega^{-1}$ cm$^{-1}$ in the moderately dirty region of $3 \times 10^3$ $\Omega^{-1}$ cm$^{-1}$ ≤ $\sigma_{xx}$ ≤ $5 \times 10^5$ $\Omega^{-1}$ cm$^{-1}$. The present result follows the tendency calculated theoretically in [Ref.24]. Thus, we consider that the intrinsic mechanism is the dominant process for the AHE of Ni/Pt (001) superlattice.

**Figure 3(a)** depicts the measurement setup for the ANE. By heating one side of the substrate, the temperature gradient was induced along the in-plane *x* direction. *H* was applied along the out-of-plane *z* direction. As a result, the $E_{ANE}$ was detected along the *y* direction. The values of $S^{ANE}$ were measured for the Hall-cross-shaped devices with the 2.0 mm-wide channel and the 2.1 mm-wide branches. Before microfabricating the devices, the Seebeck coefficient ($S$) and the longitudinal conductivities were measured for the blanket films. We also evaluated the anomalous Hall angles using the identical Hall-cross-shaped devices. In this study, when those parameters were obtained, the whole multilayer was regarded as one ferromagnetic material. **Figure 3(b)** shows the $H$ dependence of $E_{ANE}$ divided by $\nabla T$ for $t$ = 3.0 nm. The square-shaped hysteresis of $E_{ANE}/\nabla T$, which resembles the magnetization curve (**Fig. 1(a)**), was definitely observed. $S^{ANE}$ was calculated from the slope of linear fit to the plot of $E_{ANE}$ as a function of $\nabla T$ (the inset of **Fig. 3(b)**).

**Figure 3(c)** plots the $t$ dependence of $S^{ANE}$. All the samples exhibit the large values of $S^{ANE}$ ≥ 0.9 μV K$^{-1}$, and the maximum $S^{ANE}$ = 1.14±0.05 μV K$^{-1}$ was obtained at $t$ = 2.0 nm. It is noted that these $S^{ANE}$ for the present Ni/Pt superlattice are one order of magnitude of larger than that for the bulk Ni [Refs.12, 25]. In order to elucidate the mechanism of enhanced ANE for the Ni/Pt superlattice, $S$ and $\rho_{xy}/\rho_{xx}$ referred to the AHE angle are plotted as a function of $t$ in **Figs. 3(d) and 3(e)**, respectively, where $\rho_{xy} = -\sigma_{xy}/\sigma_{xx}^2$. Using the resistivity tensor, $S^{ANE}$ is expressed as [Ref.15]

$$S^{ANE} = \rho_{xx}\alpha_{xy} + \rho_{xy}\alpha_{xx}, \quad (1)$$

where $\alpha_{xx}$ is given by $S/\rho_{xx}$. The second term of Eq. (1) comes from the Seebeck effect-induced charge current, i.e. $S\rho_{xy}/\rho_{xx}$. On the other hand, the first term of Eq. (1) expresses the contribution of direct generation of transverse charge current originating from $\alpha_{xy}$. **Figure 3(f)** shows the $t$ dependence of $\rho_{xy}\alpha_{xx}$, which is two orders of magnitude smaller than $S^{ANE}$ shown in **Fig. 3(c)**. This fact definitely indicates that the conversion process through the Seebeck effect followed by AHE hardly contributes to the ANE of the Ni/Pt superlattices. Since the AHE angle of Ni/Pt superlattices is not so small compared to other ferromagnets [Ref.25], the small $S$ is the reason for the small $\rho_{xy}\alpha_{xx}$. In contrast to $\alpha_{xx}$, $\alpha_{xy}$ is the essential parameter of the large ANE of the Ni/Pt superlattices. The values of $\alpha_{xy}$ have been estimated using the obtained parameters of $S^{ANE}$, $\rho_{xx}$, $\rho_{xy}$ and $S$. **Figures 3(g)** and **3(h)** show $\alpha_{xy}$ and $\rho_{xx}$, respectively. The Ni/Pt superlattices possess very large $\alpha_{xy}$, and the maximum value is $\alpha_{xy} = 4.8$ A K$^{-1}$ m$^{-1}$ at $t = 4.0$ nm. This $\alpha_{xy}$ is comparable to or larger than several materials exhibiting large ANE such as Co$_2$MnGa (2.4 - 3.0 A K$^{-1}$ m$^{-1}$) [Ref.9], Co$_3$Sn$_2$S$_2$ (~ 2 A K$^{-1}$ m$^{-1}$) [Ref.26], and SmCo$_5$ (4.6 A K$^{-1}$ m$^{-1}$) [Ref.12].

In addition to the finding of large $\alpha_{xy}$, another feature of Ni/Pt superlattice is the large value of $S^{ANE}$ per magnetization, *i.e.* $S^{ANE}/M_s$. As shown in **Fig. 3(i)**, $S^{ANE}/M_s$ is remarkably increased for small $t$, *e.g.* 3.6 µV K$^{-1}$ T$^{-1}$ at $t = 1.5$ nm. The sample with $t = 1.5$ nm showing the small $M_s$ still maintains large $S^{ANE}$. This means interestingly that the value of $S^{ANE}$ is not proportional to the magnitude of $M_s$ even for the identical superlattices. The present Ni/Pt superlattices do not follow the relationship between $S^{ANE}$ and $M_s$ mentioned in [Ref.23]. From the viewpoint of practical applications, this large $S^{ANE}$ and small $M_s$ could be promising to improve the thermoelectric conversion performance as discussed in Refs.14, 23 and 27.

One may think the following scenarios for explaining the enhanced ANE: (i) alloying of Ni and Pt at the interfaces, (ii) the contribution of proximity-induced magnetic moments in Pt [Refs.28,29], and (iii) the enhancement of spin-orbit interaction at the interfaces. All these possibilities originate from the interface. Since the decrease in $t$ at the fixed total thickness ($t_{total}$) means the increase in the interface density, $S^{ANE}$ should increase with reducing $t$ if these interface effects are dominant. However, $t$ dependence of $S^{ANE}$ does not show remarkable increase at the small $t$. Thus, we need to consider another possible contribution. For this purpose, the first-principles calculations were made for $\sigma_{xy}$ and $\alpha_{xy}$. As described above, the large $\alpha_{xy}$ leads to the enhanced ANE. $\alpha_{xy}$ is expressed as [Ref.30]

$$\alpha_{xy} = -\frac{\pi^2}{3}\frac{k_B^2 T}{e}\left(\frac{\partial \sigma_{xy}}{\partial \varepsilon}\right)_{E_F}, \qquad (2)$$

where $k_B$ is the Boltzmann constant and $e$ is the elementary charge of electron. $\left(\partial \sigma_{xy}/\partial \varepsilon\right)_{E_F}$ is the energy derivative of $\sigma_{xy}$ at the Fermi level ($E_F$). The density-functional theory (DFT) with the aid of the Vienna *ab initio* simulation program (VASP) was used for the calculations [Ref.31]. For the details, see the Supplementary Material. The generalized gradient approximation (GGA) was adopted for the exchange-correlation energy [Ref.32], and the projector augmented wave pseudopotential [Refs.33,34] was used to treat the core electrons properly. **Figure 4** shows the $\sigma_{xy}$ and $\alpha_{xy}$ versus chemical potential ($\mu$) for the Ni 14 monolayer (ML)/Pt$d_{Pt}$ ML (Ni14/Pt$d_{Pt}$), where $d_{Pt}$ was set at 0, 2, 4 and 6, the Ni 14 ML with the vacuum interface (Ni14/vac), and the bulk Ni. Here, $\mu = 0$ corresponds to $E_F$. The in-plane lattice constants were set to 0.372 nm for the Ni 14 ML, which was determined from the experimental value [Ref.22], and 0.352 nm for the bulk Ni. For the present calculation, the Coulomb interaction ($U$) of 3.9 eV and the Hund coupling ($J$) of 1.1 eV are considered in Ni 3$d$ states as well as Ref.21. As shown in **Fig. 4(a)**, the bulk Ni and Ni14/Pt0 exhibit similar $\mu$ dependence. Note here that Ni14Pt0 does not include the vacuum layer, and a small difference between the bulk Ni and Ni14/Pt0 comes from the difference in the in-plane lattice constants. However, a drastic change is observed for Ni14/vac (**Fig. 4(a)**) and Ni14/Pt$d_{Pt}$ with $d_{Pt}$ = 2, 4 and 6 (**Fig. 4(b)**). Fine oscillatory behavior is seen in $\sigma_{xy}$ versus $\mu$. Because this feature is not observed for bulk Ni, the oscillation is attributable to the formation of interface. This oscillation in $\sigma_{xy}$ against $\mu$ leads to the increase in energy derivative of $\sigma_{xy}$. Let us here remember that the large derivative of $\sigma_{xy}$ yields a large $\alpha_{xy}$ following Eq. (2). As a result, the larger $|\alpha_{xy}|$ than that of bulk Ni was obtained at many $\mu$ values for Ni14/vac (**Fig. 4(c)**) and Ni14/Pt$d_{Pt}$ with $d_{Pt}$ = 2, 4 and 6 (**Fig. 4(d)**). The band structures and the Berry curvatures were also calculated in Ni14/Pt6 and Ni14/Pt0 (see Fig. S1 of the Supplementary Material). In the case of Ni14/Pt6 [Fig. S1(c)], the band-folding effect provides many band dispersions around $E_F$ in the ($k_x$,$k_y$) plane (corresponding to in-plane wave vectors) and the hybridizations of these bands lead to many band splittings. This is the origin for the oscillation in the Berry curvature [Fig. S1(d)], and the resultant oscillatory behavior of $\sigma_{xy}$. From these results, we may say that the oscillatory behavior in $\sigma_{xy}$ due to the interface formation is related with the enhanced ANE. Although this is another possible scenario, the present calculation also cannot fully explain the $t$ dependence of $S^{ANE}$ because the amplitude and energy position of oscillation in $\alpha_{xy}$ does not simply vary with $t_{Pt}$. For quantitative comparison and more concrete examination, further systematic studies with other materials systems including the effects of structural imperfections and/or phonon/magnon excitations are required.

Hereafter, let us discuss the contribution of interface between the Ni initial layer and the SrTiO$_3$ substrate to the ANE signal. The high-density two-dimensional electron gas confined at the interface with SrTiO$_3$ is famous for its large Seebeck coefficient [Ref.35]. If an oxygen-deficient layer and/or a Ni-doped layer exists at the interface and becomes conductive, they may affect the ANE signals. For investigating the influence of the SrTiO$_3$ substrate, we also prepared the [Ni (3.0 nm)/Pt (1.0 nm)]$_{\times N}$ superlattices with different repetition numbers: $N$ = 3, 5, 10, and 20. The different $N$ leads to the different $t_{total}$ of Ni/Pt superlattice. **Table 1** summarizes $S^{ANE}$ for the samples with different $N$ ($t_{total}$). There is no remarkable difference in $S^{ANE}$ between $N$ = 5, $N$ = 10 and $N$ = 20. However, a definite increase in $S^{ANE}$ is seen for $N$ = 3, suggesting the possibility that the ANE signal originating from the interface with the SrTiO$_3$ substrate is included in the $S^{\text{ANE}}$ for small $N$. Judging from this, we consider that the contribution from the SrTiO$_3$ substrate is negligibly small for the samples having $t_{\text{total}} \geq 20$ nm.

Finally, we quantitatively compare $S^{\text{ANE}}$ for the Ni/Pt superlattices to those for bulk Ni and a Ni single layer film. According to Ref. 12, the bulk Ni showed $S^{\text{ANE}} = 0.22$ µV K$^{-1}$. As a reference sample, we also prepared a 20 nm-thick Ni single layer on a SrTiO$_3$ substrate, which showed $S^{\text{ANE}} = 0.52 \pm 0.05$ µV K$^{-1}$. Those values are smaller than the maximum $S^{\text{ANE}}$ for the Ni/Pt superlattices. One may think that the large transverse Peltier coefficient (2.4 A K$^{-1}$ m$^{-1}$ ≤ $\alpha_{xy}$ ≤ 4.8 A K$^{-1}$ m$^{-1}$) is the key parameter for enhancing ANE in Ni/Pt superlattices. However, $\alpha_{xy}$ for bulk Ni is as large as 2.6 A K$^{-1}$ m$^{-1}$ [Ref.25]. Nevertheless, $S^{\text{ANE}}$ for bulk Ni is one order of magnitude smaller than that for the Ni/Pt superlattices, which results from the small $\rho_{xx}$ (~ 9 µΩ cm) for bulk Ni. This fact suggests that the formation of superlattice allows us to control several key parameters independently thanks to the degrees of freedom in design that the superlattice structure possesses.

In summary, we demonstrated the enhancement of ANE owing to the formation of Ni/Pt superlattices. The value of $S^{\text{ANE}}$ was increased up to more than 1 µV K$^{-1}$ for the samples with 2.0 nm ≤ $t$ ≤ 4.0 nm, and the large $S^{\text{ANE}}/M_s = 3.6$ µV K$^{-1}$ T$^{-1}$ was achieved for $t = 1.5$ nm. The enhanced ANE is attributable to the large $\alpha_{xy}$. We believe that the present study on Ni/Pt superlattices provides a strategy to enhance ANE.

**Acknowledgement**

The authors thank K. Ito and H. Kurebayashi for their valuable comments and M. Isomura for technical supports. The device fabrication was partly carried out at the Cooperative Research and Development Center for Advanced Materials, IMR, Tohoku University. This work was supported by the Grant-in-Aid for Scientific Research (S) (JP18H05246) from JSPS KAKENHI, PRESTO from the Japan Science and Technology Agency (No. JPMJPR17R5), and CREST "Creation of Innovative Core Technologies for Nano-enabled Thermal Management" (JPMJCR17I1) from JST, Japan. A.M. is supported by JSPS through a research fellowship for young scientists (JP18J02115).

**References**

[1]  G. E. W. Bauer, E. Saitoh, and B. J. van Wees, *Nat. Mater.* **11**, 391 (2012).

[2]  K. Uchida, S. Takahashi, K. Harii, J. Ieda, W. Koshibae, K. Ando, S. Maekawa, and E. Saitoh, *Nature* **455**, 778 (2008).

[3]  K. Uchida, J. Xiao, H. Adachi, J. Ohe, S. Takahashi, J. Ieda, T. Ota, Y. Kajiwara, H. Umezawa, H. Kawai, G. E. W. Bauer, S. Maekawa, and E. Saitoh, *Nat. Mater.* **9**, 894 (2010).

[4] C. M. Jaworski, J. Yang, S. Mack, D. D. Awschalom, J. P. Heremans, and R. C. Myers, *Nat. Mater.* **9**, 898 (2010).

[5] Y. Sakuraba, *Scri. Mater.* **111**, 29 (2016).

[6] Y. Sakuraba, K. Hasegawa, M. Mizuguchi, T. Kubota, S. Mizukami, T. Miyazaki, K. Takanashi, *Appl. Phys. Exp.* **6**, 033003 (2013).

[7] T. Seki, R. Iguchi, K. Takanashi, and K. Uchida, *J. Phys. D: Appl. Phys.*, 51, 254001 (2018).

[8] K. Hasegawa, M. Mizuguchi, Y. Sakuraba, T. Kamada, T. Kojima, T.Kubota, S. Mizukami, T. Miyazaki, and K. Takanashi, *Appl. Phys. Lett.* **106**, 252405 (2015).

[9] A. Sakai, Y. P. Mizuta, A. A. Nugroho, R. Sihombing, T. Koretsune, M. Suzuki, N. Takemori, R. Ishii, D. Nishio-Hamane, R. Arita, P. Goswami and S. Nakatsuji, *Nat. Phys.* **14**, 1119 (2018).

[10] T. Seki, R. Iguchi, K. Takanashi, and K. Uchida, *Appl. Phys. Lett.* **112**, 152403 (2018).

[11] H. Reichlova, R. Schlitz, S. Beckert, P. Swekis, A. Markou, Y. Chen, D. Kriegner, S. Fabretti, G. H. Park, A. Niemann, S. Sudheendra, A. Thomas, K. Nielsch , C. Felser, and S. T. B. Goennenwein, *Appl. Phys. Lett.* **113**, 212405 (2018).

[12] A. Miura, H. Sepehri-Amin, K. Masuda, H. Tsuchiura, Y. Miura, R. Iguchi, Y. Sakuraba, J. Shiomi, K. Hono, and K. Uchida, *Appl. Phys. Lett.* **115**, 222403 (2019).

[13] H. Nakayama, K. Masuda, J. Wang, A. Miura, K. Uchida, M. Murata, and Y. Sakuraba, *Phys. Rev. Mater.* **3**,

114412 (2019).

[14] W. Zhou and Y. Sakuraba, *Appl. Phys. Exp.* **13**, 043001 (2020).

[15] Y. Sakuraba, K. Hyodo, A. Sakuma, and S. Mitani, *Phys. Rev. B* **101**, 134407 (2020).

[16] A. Miura, K. Masuda, T. Hirai, R. Iguchi, T. Seki, Y. Miura, H. Tsuchiura, K. Takanashi and K. Uchida, *Appl. Phys. Lett.* **117**, 082408 (2020).

[17] K. Uchida, T. Kikkawa, T. Seki, T. Oyake, J. Shiomi, Z. Qiu, K. Takanashi and E. Saitoh, *Phys. Rev. B* **92**, 094414 (2015).

[18] C. Fang, C. H. Wan, Z. H. Yuan, L. Huang, X. Zhang, H. Wu, Q. T. Zhang, and X. F. Han, *Phys. Rev. B* **93**, 054420 (2016).

[19] K. Uchida, S. Daimon, R. Iguchi, and E. Saitoh, *Nature* **558**, 95 (2018).

[20] K. Masuda, K. Uchida, R. Iguchi, and Y. Miura, *Phys. Rev. B* **99**, 104406 (2019).

[21] J. Weischenberg, F. Freimuth, S. Blügel, and Y. Mokrousov, *Phys. Rev. B* **87**, 060406 (2013).

[22] T. Seki, M. Tsujikawa, K. Ito, K. Uchida, H. Kurebayashi, M. Shirai, and K. Takanashi, *Phys. Rev. Mater.* **4**, 064413-1-9 (2020).

[23] M. Ikhlas, T. Tomita, T. Koretsune, M. Suzuki, D. Nishio-Hamane, R. Arita, Y. Otani and S. Nakatsuji, *Nat. Phys.* **13**, 1085 (2017).

[24] S. Onoda, N. Sugimoto, and N. Nagaosa, *Phys. Rev. B* **77**, 165103 (2008).

[25] A. Miura, R. Iguchi, T. Seki, K. Takanashi, and K. Uchida, *Phys. Rev. Mater.* **4**, 034409 (2020).

[26] S. N. Guin, P. Vir, Y. Zhang, N. Kumar, S. J. Watzman, C. Fu, E. Liu, K. Manna, W. Schnelle, J. Gooth, C. Shekhar, Y. Sun, and C. Felser, *Adv. Mater.* **31**, 1806622 (2019).

[27] T. Seki, A. Miura, K. Uchida, T. Kubota, and K. Takanashi, *Appl. Phys. Exp.* **12**, 023006 (2019).

[28] T. Kikkawa, M. Suzuki, R. Ramos, M. H. Aguirre, J. Okabayashi, K. Uchida, I. Lucas, A. Anadón, D. Kikuchi, P. A. Algarabel, L. Morellón, M. R. Ibarra, and E. Saitoh, *J. Appl. Phys.* **126**, 143903 (2019).

[29] C. Klewe, T. Kuschel, J.-M. Schmalhorst, F. Bertram, O. Kuschel, J. Wollschläger, J. Strempfer, M. Meinert, and G. Reiss, *Phys. Rev. B* **93**, 214440 (2016).

[30] N. F. Mott and H. Jones, in Clarendon Press. Oxford (1936), pp. 308–314.

[31] G. Kresse and J. Furthmüller, *Phys. Rev. B* **54**, 11169 (1996).

[32] J. P. Perdew, K. Burke, and M. Ernzerhof, *Phys. Rev. Lett.* **77**, 3865(1996).

[33] P. E. Blöchl, *Phys. Rev. B* **50**, 17953 (1994).

[34] G. Kresse and D. Joubert, *Phys. Rev. B* **59**, 1758 (1999).

[35] H. Ohta, S. Kim, Y. Mune, T. Mizoguchi, K. Nomura, S. Ohta, T. Nomura, Y. Nakanishi, Y. Ikuhara, M. Hirano, H. Hosono, and K. Koumoto, *Nat. Mater.* **6**, 129 (2007).

**Table 1**  Anomalous Nernst coefficient ($S^{\text{ANE}}$) for [Ni (3.0 nm)/Pt (1.0 nm)]$_{\times N}$ superlattices with different repetitions.

| | $N = 3$ | $N = 5$ | $N = 10$ | $N = 20$ |
|---|---|---|---|---|
| $t_{\text{total}}$ (nm) | 12.0 | 20.0 | 40.0 | 80.0 |
| $S^{\text{ANE}}$ (μV K$^{-1}$) | 1.49 ± 0.17 | 1.13 ± 0.17 | 1.14 ± 0.18 | 1.17 ± 0.05 |

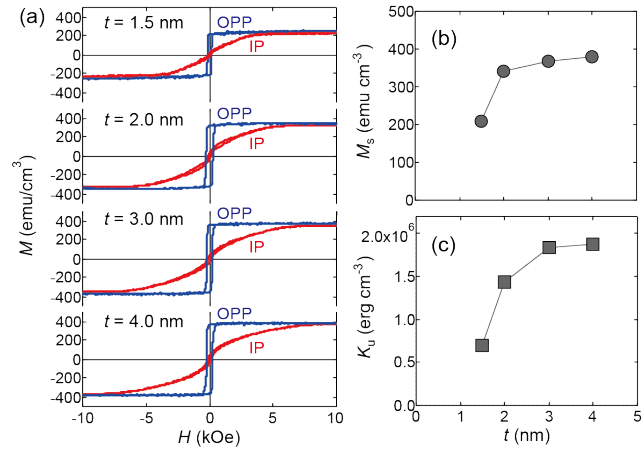

**Figure 1** (a) Magnetization curves for the [Ni ($t$)/Pt (1.0 nm)] $_{\times N}$ with $t$ = 1.5, 2.0, 3.0 and 4.0 nm, where $N$ was set to be 8, 7, 5, and 4, respectively. The red curves denote the magnetization curves measured with the magnetic field ($H$) applied in the film plane (IP curve) while the blue curves denote those measured with out-of-plane magnetic field (OPP curve). (b) Saturation magnetization ($M_s$) and uniaxial magnetic anisotropy constant ($K_u$) as a function of $t$.

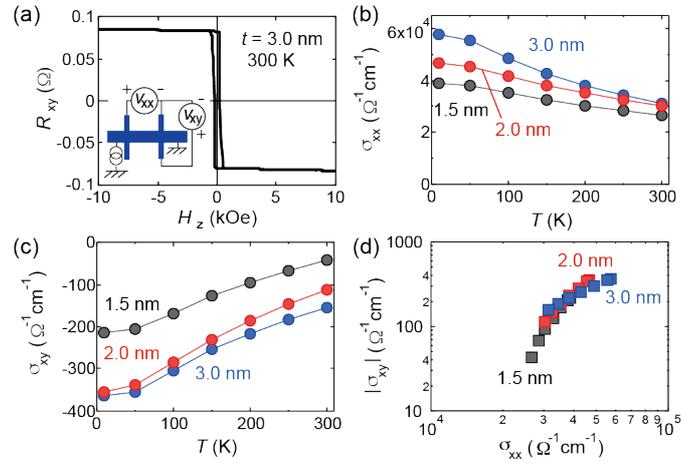

**Figure 2** (a) Transverse resistance ($R_{xy}$) as a function of perpendicular magnetic field ($H_z$) for the device with $t = 3.0$ nm. Inset: the illustration of Hall device structure and measurement setup. (b) Longitudinal conductivity ($\sigma_{xx}$) and (c) transverse conductivity ($\sigma_{xy}$), at $H_z = 0$ Oe as a function of measurement temperature ($T$). (d) $\sigma_{xx}$ versus $|\sigma_{xy}|$ plot.

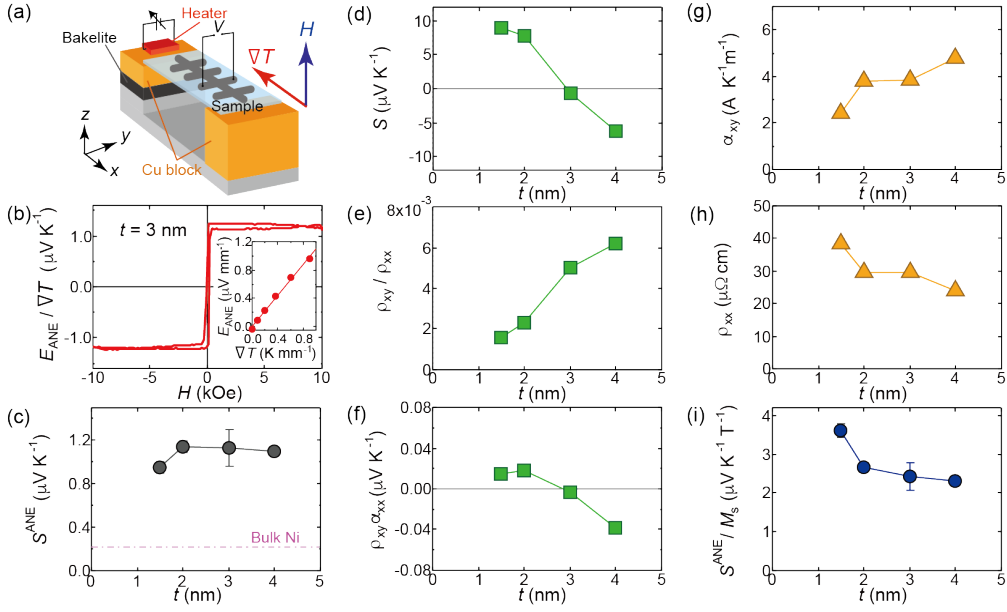

**Figure 3** (a) Measurement setup for anomalous Nernst effect (ANE). (b) $H$ dependence of electric field induced by ANE ($E_{ANE}$) divided by $\nabla T$ for the devices with $t$ = 3.0 nm. Inset: the plot of $E_{ANE}$ as a function of $\nabla T$. (c) $t$ dependence of anomalous Nernst coefficient ($S^{ANE}$), (d) Seebeck coefficient ($S$), (e) $\rho_{xy}/\rho_{xx}$, (f) $\rho_{xy}\alpha_{xx}$, (g) transverse Peltier coefficient ($\alpha_{xy}$), (h) $\rho_{xx}$ and (i) $S^{ANE}/M_s$. In (c), the dashed-dotted line denotes the value for the bulk Ni reported in Ref. 12.

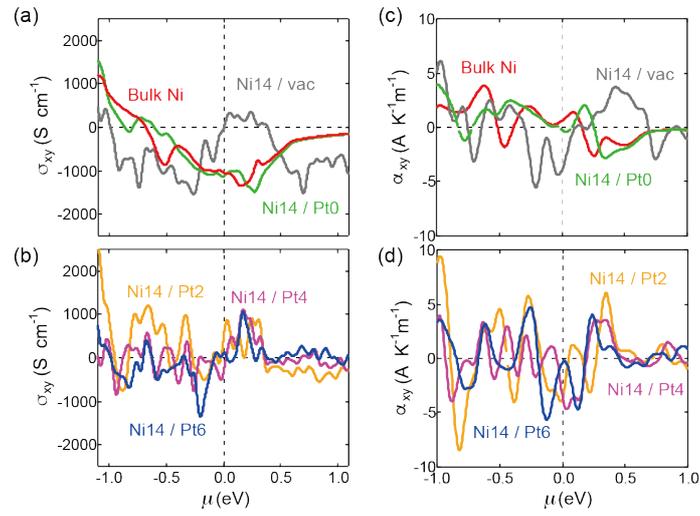

**Figure 4** First-principles calculation of $\sigma_{xy}$ ((a) and (b)) and $\alpha_{xy}$ ((c) and (d)) versus chemical potential ($\mu$) for the Ni 14 monolayer (ML)/Pt$d_{Pt}$ML, where $d_{Pt}$ was set at 0, 2, 4 and 6, the Ni 14 ML with the vacuum interface, and the bulk Ni. The calculations for $\alpha_{xy}$ were carried out at 300 K.